\documentclass[%
aps,
prl,
 amsmath,amssymb,
reprint,
]{revtex4-1}

\usepackage{graphicx}
\usepackage{dcolumn}
\usepackage{bm}
\usepackage{color}

\usepackage{amsmath}	
\begin{document}

\newcommand{\diff}[2]{\frac{d#1}{d#2}}
\newcommand{\pdiff}[2]{\frac{\partial #1}{\partial #2}}
\newcommand{\fdiff}[2]{\frac{\delta #1}{\delta #2}}
\newcommand{\bx}{\bm{x}}
\newcommand{\bv}{\bm{v}}
\newcommand{\br}{\bm{r}}
\newcommand{\ba}{\bm{a}}
\newcommand{\by}{\bm{y}}
\newcommand{\bY}{\bm{Y}}
\newcommand{\bF}{\bm{F}}
\newcommand{\bn}{\bm{n}}
\newcommand{\be}{\bm{e}}
\newcommand{\new}{\nonumber\\}
\newcommand{\abs}[1]{\left|#1\right|}
\newcommand{\tr}{{\rm Tr}}
\newcommand{\HH}{{\mathcal H}}
\newcommand{\OO}{{\mathcal O}}
\newcommand{\var}{{\rm Var}}
\newcommand{\ave}[1]{\left\langle #1 \right\rangle}

\preprint{AIP/123-QED}

\title{ Comment on ``A new universality class describes Vicsek's
flocking phase in physical dimensions''}
\author{Harukuni Ikeda}
 \email{harukuni.ikeda@gakushuin.ac.jp}
\affiliation{
Department of Physics, Gakushuin University, 1-5-1 Mejiro, Toshima-ku, Tokyo 171-8588, Japan}


\date{\today}




\maketitle


\section{Introduction}

In a recent preprint, ``A new universality class describes Vicsek's
flocking phase in physical dimensions'', Patrick Jentsch and Chiu Fan
Lee have computed the critical exponents of the Vicsek model in the
ordered phase by means of functional renormalization group
methods~\cite{jentsch2024}. In this note, we compare their results with
our previous theoretical predictions for the Vicsek model, which is
expected to be exact in $d=2$ dimensions~\cite{ikeda2024advection}. We
point out that the critical exponents predicted by the two theories are
extremely close in both $d=2$ and $3$ dimensions. We found that both theories
fit the current numerical data equally well. Extensive numerical simulations
for larger system sizes are thus highly desirable to judge which
theory is correct.


\section{Exact results in $d=2$}
Here, we first briefly explain our theoretical
prediction~\cite{ikeda2024advection}. The Vicsek model consists of $XY$
spins that fly along their magnetic directions~\cite{vicsek1995}. In the
ordered phase, the rotational symmetry is broken spontaneously, and thus,
the Nambu-Goldstone (NG) modes arise~\cite{nambu1960,goldstone1962}.  In
a previous work, we have derived the critical exponents that
characterize the scaling behavior of the NG
modes~\cite{ikeda2024advection}. The calculation is actually quite
straightforward in $d=2$. To make this note self-consistent, we here
provide a refined derivation of our previous theoretical prediction in
$d=2$~\cite{ikeda2024advection}.

Let $\bv(\bx)$ be the coarse-grained velocity field of spins at around
$\bx$.  In the ordered phase, the spins are aligned in some direction
$x_\parallel$, and the average velocity has a finite value
$\ave{\bv}=v_{0}\be_\parallel$, where $\be_\parallel$ denotes the unit
vector along $x_\parallel$.  There is a zero mode
corresponding to the global rotation, which does not change the norm of
$\bv$. In $d=2$, this rotational motion can be represented by 
the angle $\theta$ of the velocity field,
$\bv=\{v_\parallel,v_\perp\}=\{v_0\cos\theta,v_0\sin\theta\}$. The NG
modes are often recognized as the modes associated with $v_\perp$.
However, strictly speaking, that is correct only for the linear order
$v_\perp\approx v_0\theta$~\cite{toner1995,ikeda2024advection}, and it
is better to use $\theta$ to represent the NG
modes~\cite{sartori2019}. We assume that the EOM of the
NG modes, $\theta$, is written as a closed form of $\theta$:
\begin{align}
\dot{\theta}(x_\parallel,x_\perp,t) = F[\theta] + \xi(x_\parallel,x_\perp,t),
\end{align}
where $F$ denotes the restitution force of the NG mode, and $\xi$ denotes the white
noise of zero mean and variance:
\begin{align}
\ave{\xi(x_\parallel,x_\perp,t)\xi(x_\parallel',x_\perp',t')}=
 2D\delta(x_\parallel-x_\parallel')\delta(x_\perp-x_\perp')\delta(t-t').
 \label{142115_7Feb24}
\end{align}
To investigate the large spatio-temporal behavior, we consider the
following scaling transformation~\cite{toner1995,toner1998,sartori2019}:
\begin{align}
x_\perp\to bx_\perp,\ x_\parallel\to b^{\zeta}x_\parallel,\  t\to b^{z}t,\ \theta\to b^{\chi}\theta.
\end{align}
From the naive dimensional analysis, one can infer the scaling
dimensions as $\dot{\theta}\sim b^{\chi-z}$.  Also,
Eq.~(\ref{142115_7Feb24}) implies $\ave{\xi^2} \sim b^{-z-1-\zeta}$, leading
to $\xi\sim b^{-(z+1+\zeta)/2}$. Requiring $\dot{\theta}\sim \xi$, we
get the hyper-scaling relation
\begin{align}
z = 2\chi + \zeta + 1,\label{142611_7Feb24}
\end{align}
which is a special case of the hyper-scaling reported in a previous
numerical simulation of the Vicsek model in $d=2$ and $3$
dimensions~\cite{mahault2019}
\begin{align}
z = 2\chi + \zeta + d-1.\label{145516_7Feb24}
\end{align}
In general, the naive scaling analysis does not give the
correct results for non-linear systems~\cite{nishimori2011elements}.
However, as we will see later, the scaling relation Eq.~(\ref{142611_7Feb24})
turns out to be exact in $d=2$. 

Above the lower critical dimension, the exponent $\chi$ should be
negative $\chi<0$, meaning that $\theta\sim b^{\chi}\to 0$ in the
thermodynamic limit $b\to\infty$~\cite{toner1995}. Also,
$\partial_\parallel \sim b^{-\zeta}\to 0$ and $\partial_\perp\sim
b^{-1}\to 0$ in the limit $b\to \infty$. Therefore, $F[\theta]$ can be
expanded by $\theta$, $\partial_\parallel$, and $\partial_\perp$ as
\begin{align}
F[\theta] = f_1(\partial_\parallel,\partial_\perp)\theta +
 f_2(\partial_\parallel,\partial_\perp)\theta^2
+O(\partial_{\parallel}\theta^3, \partial_\perp \theta^3),
\end{align}
where $f_1$ and $f_2$ are some unknown functions, and higher-order terms
are negligible since they have smaller scaling dimensions.  The linear 
order term $f_1\theta$ is not necessary to calculate the
critical exponents, as we will see below. The second-order term would be
expanded as
\begin{align}
f_2(\partial_\parallel,\partial_\perp)\theta^2 = (a\partial_\parallel +b\partial_\perp)\theta^2
 + O(\partial_\parallel^2\theta^2,\partial_\perp^2\theta^2),
\end{align}
where $a$ and $b$ are some constants. Then, the EOM in $d=2$ reduces to
\begin{align}
\dot{\theta} \approx 
({\rm linear })
 + (a\partial_\parallel + b\partial_\perp)\theta^2 + \xi.\label{130116_7Feb24}
\end{align}
Note that the non-linear terms are written as a total derivative of
$\partial_\parallel$ or $\partial_\perp$, which only affects the scaling
dimensions of the terms involving $\partial_\parallel$ or
$\partial_\perp$~\cite{toner1995,toner1998,sartori2019}. In particular, the scaling
dimensions of $\dot{\theta}$ and $\xi$ estimated by the naive scaling
analysis should remain unchanged~\cite{mahault2019}. Therefore, the hyperscaling
Eq.~(\ref{142611_7Feb24}) holds exactly in $d=2$. Also, one can see that
Eq.~(\ref{130116_7Feb24}) is invariant under the ‘pseudo-Galilean’
transformations: $\theta\to \theta+\Theta$, $x_\parallel\to x_\parallel+2a\Theta t$, and
$x_\perp\to x_\perp+2b\Theta t$, which implies the following scaling
relations~\cite{toner1995,sartori2019}:
\begin{align}
&1 = \chi+z,\label{143047_7Feb24}\\
&\zeta = \chi + z.\label{143053_7Feb24}
\end{align}
Using Eq.~(\ref{142611_7Feb24}), (\ref{143047_7Feb24}), and (\ref{143053_7Feb24}),
one can determine the exact critical exponents in $d=2$~\cite{ikeda2024advection}:
\begin{align}
 \chi = -\frac{1}{3},\ z = \frac{4}{3},\ \zeta = 1.\label{084027_8Feb24}
\end{align}
Since Eq.~(\ref{130116_7Feb24}) is the most general form of the EOM, we
conclude that Eqs.~(\ref{084027_8Feb24}) also give the exact critical
exponents for the Vicsek model~\cite{ikeda2024advection}.

For $d>2$, it is not clear if the system satisfies the hyper-scaling
Eq.~(\ref{145516_7Feb24}) and pseudo-Galilean invariance
Eqs.~(\ref{143047_7Feb24}) and (\ref{143053_7Feb24}). Therefore, we
could not determine the exact critical exponents.  Instead, by means of
a scaling argument of the Toner-Tu hydrodynamic
theory~\cite{toner1995,toner1998,toner2012}, we derived approximated values of the critical
exponents~\cite{ikeda2024advection}:
\begin{align}
\chi = \frac{1-d}{3},\ z = \frac{d+2}{3},\ \zeta = 1.
\end{align}
Note that the same result can also be obtained by assuming that
Eqs.~(\ref{145516_7Feb24}), (\ref{143047_7Feb24}), and
(\ref{143053_7Feb24}) hold in $d>2$.  In Table \ref{235502_31Dec23}, we
compare our theoretical prediction (Ikeda24) with the results of the
recent numerical simulation of the Vicsek model~\cite{mahault2019}. The
agreement is reasonably good.

\begin{table}[t]
\begin{center}
\caption{Critical exponents.}  \label{235502_31Dec23}
\begin{tabular}{l|rrr|rrr}
&$d = 2$ &  & & $d=3$ & & \\ 
      & Simulation& Ikeda24&JL24& Simulation& Ikeda24&JL24\\ \hline
$\chi$&-0.31(2)&   -0.333&-0.325    &-0.62         &  -0.667    &-0.65\\
$z$& 1.33(2)   &   1.333&1.325      &1.77       &1.667    &1.65\\
$\zeta$&0.95(2)&     1   &0.975     &1       &1      &0.95\\
\end{tabular}
\end{center}
\end{table}

\section{Functional renormalization group}

\begin{figure}[t]
\includegraphics[width=8cm]{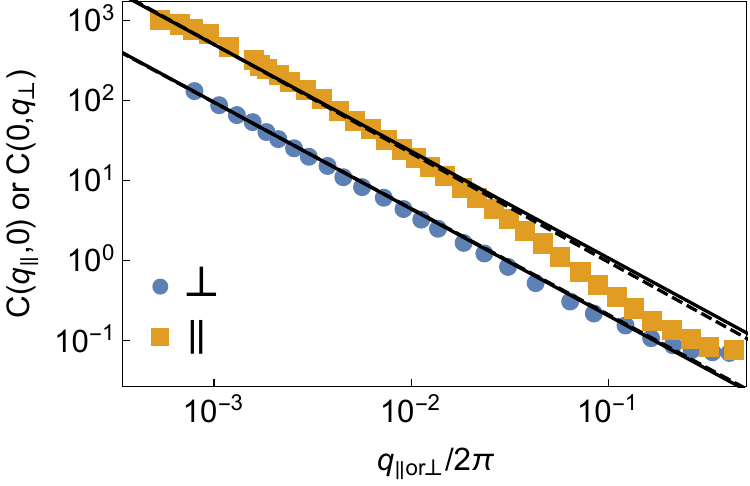} \caption{Correlation
 functions. Markers denote the numerical results of the
 Vicsek model in $d=2$ for $L=8000$, taken from Fig.1~(a) in
 Ref.~\cite{mahault2019}. The solid lines represent the fit with 
our exact results of the critical exponents, while the dashed lines represent the fit with
the results of the renormalization group calculations.}
				      \label{083952_8Feb24}
\begin{center}
\end{center}
\end{figure}

In a more recent preprint, Jentsch and Lee calculated the critical
exponents of the Vicsek model in the ordered phase by means of
functional renormalization group methods~\cite{jentsch2024}.  The
calculations are highly cumbersome, even in $d=2$.  So we here just
summarize their main results for the critical
exponents~\cite{jentsch2024}:
\begin{align}
\chi = \frac{13(1-d)}{40},\ 
z = \frac{27+13d}{40},\ \zeta = \frac{41-d}{40}.\label{221330_7Feb24} 
\end{align}
Interestingly, the above critical exponents satisfy the hyper-scaling
Eq.~(\ref{145516_7Feb24}), and first scaling relation obtained by the
'pseudo-Galielan' invariance Eq.~(\ref{143047_7Feb24}).  However,
the exponents do not satisfy the last scaling relation
Eq.~(\ref{143053_7Feb24}). As a consequence, Eq.~(\ref{221330_7Feb24})
predicts a weak anisotropic scaling behavior $\zeta<1$: $\zeta=0.975$ in $d=2$ and
$\zeta=0.95$ in $d=3$. It is important for future work to see if more
sophisticated renormalization group methods can reproduce the exact
result $\zeta=1$ in $d=2$.

\section{Comparison with numerical results}

In Table~\ref{235502_31Dec23}, we compare the result by Jentsch and Lee
(JL24) with our theory (Ikeda24)~\cite{ikeda2024advection} and numerical
results of the Vicsek model~\cite{mahault2019}. The three results are
extremely close. Another important physical quantity to compare with
numerical simulation is the correction function. In $d=2$, the scaling
behaviors $v_\perp \sim \theta \sim b^{\chi}$, $x_\perp\sim b$, and
$x_\parallel \sim b^{\zeta}$ imply~\cite{toner1995,toner1998}
\begin{align}
C(x_\parallel,x_\perp)=\ave{v_\perp(x_\parallel,x_\perp)v_\perp(0)} \sim b^{2\chi}\sim x_\perp^{2\chi}
 \sim x_\perp^{2\chi/\zeta}.
\end{align}
In the Fourier space, it becomes~\cite{toner1995,toner1998}
\begin{align}
C(q_\parallel,q_\perp) = \int d\bx e^{i\bm{q}\cdot\bx}C(\bx)
\sim b^{2\chi+\zeta+1}\sim q_\perp^{-z}\sim q_\parallel^{-z/\zeta}.
\end{align}
To calculate $C(q_\parallel,q_\perp)$ precisely, one should perform the numerical
simulation in a sufficiently large linear box size $L$ because the minimal
wave vector $q_*$ scales as $q_*\sim 2\pi/L$.  The best numerical result for
the Vicsek model has been obtained in $d=2$ for $L=8000$ in
Ref.~\cite{mahault2019}. In Fig.~\ref{083952_8Feb24}, we compare the
numerical results, our exact result in $d=2$ Eqs.~(\ref{084027_8Feb24}),
and results of the renormalization group methods
Eqs.~(\ref{221330_7Feb24}). Both theoretical predictions fit the
numerical results equally well.  With the current numerical data, it is difficult to
judge which theory is correct. Numerical simulations for larger
system sizes are highly desired.


\acknowledgments
This work was supported by KAKENHI
23K13031.



\bibliography{reference}

\end{document}